# Transient Spectroscopy of Glass-Embedded Perovskite Quantum Dots: Novel Structures in an Old Wrapping

Oleg V. Kozlov[1], Rohan Singh[1], Bing Ai[2], Jihong Zhang[2], Chao Liu[2], Victor I. Klimov[1]*

[1]Chemistry Division, C-PCS, Los Alamos National Laboratory, Los Alamos, New Mexico 87545, USA

[2]State Key Laboratory of Silicate Materials for Architectures, Wuhan University of Technology, Hubei 430070, P. R. China

*Corresponding author, e-mail: klimov@lanl.gov

**Abstract.** Semiconductor doped glasses had been used by the research and engineering communities as color filters or saturable absorbers well before it was realized that their optical properties were defined by tiny specs of semiconductor matter known presently as quantum dots (QDs). Nowadays, the preferred type of QD samples are colloidal particles typically fabricated via organometallic chemical routines that allow for exquisite control of QD morphology, composition and surface properties. However, there is still a number of applications that would benefit from the availability of high-quality glass-based QD samples. These prospective applications include fiber optics, optically pumped lasers and amplifiers, and luminescent solar concentrators (LSCs). In addition to being perfect optical materials, glass matrices could help enhance stability of QDs by isolating them from the environment and improving heat exchange with the outside medium. Here we conduct optical studies of a new type of all-inorganic $CsPbBr_3$ perovskite QDs fabricated directly in glasses by high-temperature precipitation. These samples are virtually scattering free and exhibit excellent waveguiding properties which makes them well suited for applications in, for example, fiber optics and LSCs. However, the presently existing problem is their fairly low room-temperature emission quantum yields of only *ca.* 1 - 2%. Here we investigate the reasons underlying the limited emissivity of these samples by conducting transient photoluminescence (PL) and absorption measurements across a range of temperatures from 20 – 300 K. We observe that the low-temperature PL quantum yield of these samples can be as high as ~25%. However, it quickly drops (in a nearly linear fashion) with increasing temperature. Interestingly, contrary to traditional thermal quenching models, experimental observations cannot be explained in terms of a thermally activated nonradiative rate but rather suggest the existence of two distinct QD sub-ensembles of "emissive" and completely "nonemissive" particles. The temperature-induced variation in the PL efficiency is likely due to a structural transformation of the QD surfaces or interior leading to formation of extremely fast trapping sites or nonemissive phases resulting in conversion of emissive QDs into nonemissive. Thus, future efforts on improving emissivity of glass-based perovskite QD samples might focus on approaches for extending the range of stability of the low-temperature up to room temperature.

# 1 Introduction

The field of semiconductor nanocrystals or quantum dots (QDs) started almost four decades ago with observations of effects of quantum confinement in semiconductor-doped glasses [1-4]. These glass-based samples were used to gain initial insights into the fascinating physics of zero-dimensional (0D) structures including the effects of constrained dimensions on the structure of electronic states [5, 6], the role of Auger processes in 0D materials [7-10], and the physics of optical gain and lasing [11]. Nowadays, however, the NC field is dominated by colloidal samples that can be fabricated with precisely controlled sizes, shapes, and complex internal structures using modern nanochemistry techniques [12-15]. Despite this shift of emphasis to colloidal samples, there is still a considerable interest in high-quality glass-embedded QDs. Such samples, for example, would be ideally suited for applications such as window-based sunlight collectors employing ideas of semitransparent luminescent solar concentrators (LSCs) [16-18]. Glass matrices can also be used to improve the stability of QDs by protecting them from deleterious effects of the environment and improving heat exchange with the surrounding medium. In fact, encapsulation of colloidal QDs into thin $SiO_2$ glassy shells or sol-gel matrices has been frequently used to enhance QD stability for LSC [17] and lasing [19, 20] applications.

A recent addition to the QD family has been all-inorganic perovskite nanocrystals comprised of $CsPbX_3$ (X =I, Br, Cl or a mixture of these halides) [21]. These materials have shown a considerable promise as highly luminescent, spectrally tunable fluorophores for applications in light-emitting diodes (LEDs) [22-24], lasers [25-27] and LSCs [28, 29]. One deficiency of these materials is a limited photostablity, especially in the presence of oxygen and moisture [30]. A possible mitigation of this problem would be encapsulation of perovskite QDs into a glass matrix as a means to isolate them from the environment.

Recently, the "classic" methods of high-temperature precipitation of semiconductors in molten glasses have been successfully adopted for preparing glass-embedded perovskite QDs (gp-QDs) [31, 32]. High optical quality, negligible scattering and excellent waveguiding properties of these samples make them attractive for fiber optics, lasing and LSCs applications. The photophysics of these materials is still largely unexplored. However, the initial studies indicate the feasibility of realizing high (tens of percent) emission quantum yields (QYs) with this type of samples [31, 32]. The purpose of the present work is to elucidate QY-limiting factors in gp-QDs by investigating photoexcited electronic dynamics via time-resolved photoluminescence (PL) and transient absorption (TA) monitored over a range of temperatures ($T = 20 - 300$ K) and  pump fluences. These studies reveal that traditional models utilizing a thermally-activated nonradiative rate are incapable of explaining the temperature-induced variations in the PL QY.  Instead the observed behaviors suggest the existence of sub-ensembles of emissive and nonemissive dots with relative fractions dependent on the sample temperature. More subtle variations in the emission efficiency can be linked to a thermally activated trapping manifested in initial sub-ns-to-ns dynamics. Interestingly, while the observed single-exciton relaxation in glass-embedded samples is very different from that in colloidal samples, their biexciton Auger decays are virtually identically, once again confirming the insensitivity of Auger processes to QD surface properties or the identity of the embedding medium.

# 2 Experimental Section

## 2.1 Sample preparation

Borosilicate glass with nominal composition of $54B_2O_3$-$18SiO_2$-$12CaO$-$2Cs_2O$-$8PbO$-$6NaBr$ (in mol %) was prepared using a conventional melt-quenching method [31]. Starting powders ($B_2O_3$, $SiO_2$, $CaCO_3$, $Cs_2CO_3$, PbO and NaBr), were weighed and mixed, then melted in an alumina crucible at 1150 °C for 50 min. The melt was quenched by pouring onto a brass mold and quickly pressed by another brass plate. Then, the glass was annealed at 350 °C for 3 hours to remove internal stress. The annealed glass was further heat-treated at a fixed temperature varied from 490 to 520 °C for 10 hours, which resulted in formation (precipitation) of $CsPbBr_3$ QDs. Optical absorption and PL spectra of the fabricated samples are displayed in Figure S1 (see Supplementary Material, SM). Unless it is mentioned otherwise, spectroscopic studies reported in this work have been conducted for the QD samples fabricated at 520 °C. According to transmission electron microscopy (TEM) measurements, these QDs were approximately cubic and had the mean side length $d$ = 11 nm (Figure S2).

## 2.2 Optical

Steady-state absorption spectra were measured with a PerkinElmer Lambda 950 spectrometer. Steady-state PL spectra were recorded with a Horiba FluoroMax-4 spectrometer in a reflection geometry. PL QYs were measured with an integrating sphere setup. The sample was excited with a 385 nm LED (Thorlabs M385F1) inside the integrating sphere; the resulted PL was collected with an optical fiber and detected with an Ocean Optics Jaz spectrometer.

Time-resolved, temperature-dependent PL was recorded for QD samples placed in a close-cycle helium-cooled cryostat. Excitation pulses (~100 fs duration) were produced by doubling the output of a femtosecond Ti:sapphire amplifier (Coherent RegA 9000). The excitation beam diameter was ~120 μm. The PL signal was detected with a superconducting nanowire single-photon detector (SSPD); the instrumental response function was ~70 ps.

TA measurements were performed using a standard pump-probe setup based on an amplified Yb:KGW femtosecond laser (Light Conversion Pharos) operating at the 500 Hz repetition rate; pulse duration was ~150 fs. The third harmonic of the fundamental output (343 nm) was used as a pump; the pump beam diameter was ~410 μm. A small portion of the fundamental output was focused onto a sapphire plate to generate a white-light continuum, which was used as a probe. The excitation was modulated with a mechanical chopper at 250 Hz, the transmitted white-light was detected with an Avantes AvaSpec-Fast ULS1350F-USB2 spectrometer.

### 2.3 Accelerated photostability measurements

For accelerated photostability measurements, the sample was exposed to collimated 405 nm light from a high-power LED (Thorlabs M405LP1-C1); the incident power density was ~760 mW cm$^{-2}$. Due to extremely strong absorption at the excitation wavelength (optical density OD > 3), the incident LED light was completely absorbed. If the sample were exposed to the AM1.5G solar radiation, it would absorb ~25 mW cm$^{-2}$. Thus, the acceleration factor was ~30. Taking into account that in standard outdoor conditions the sample would be exposed to sunlight approximately half of the day on average, the actual acceleration factor is twice as large, i.e. ~60. We use this value to translate the laboratory time into the effective time in outdoor conditions.

## 3 Results and discussion

### 3.1 Temperature-dependent emission spectra and emission quantum yields

In Figure 1a, we display room-temperature ($T$ = 300 K) absorption (black line) and PL (red line) spectra of gp-QDs fabricated at 520 °C (Figure S1 of SM shows optical spectra of samples fabricated at other temperatures). The absorption spectrum shows a sharp onset at ~2.35 eV, but without a pronounced band-edge peak. In order to determine the energy of the lowest-energy QD optical transition, which defines the QD band gap ($E_g$), we calculate the second derivative of the absorption spectrum (blue line in Figure 1a) and find the position of its minimum. Based on this procedure, $E_g$ = 2.433 eV. The PL (Figure 1a, red line) is centered around 2.35 eV and has a fairly narrow linewidth of ~90 meV (full width at half maximum, FWHM). These spectroscopic parameters of glass-based samples are close to those of previously reported cubically shaped colloidal CsPbBr$_3$ QDs with the side length of 10.7 nm [33].

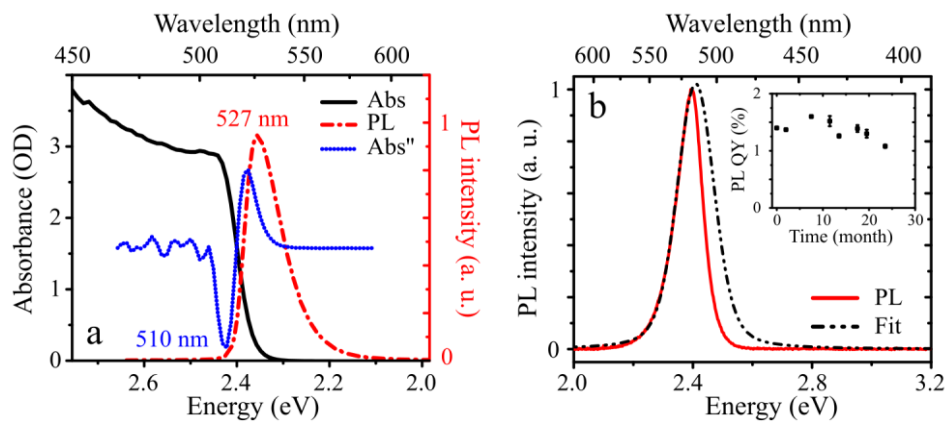

**Figure 1:** Absorption and photoluminescence spectra. (a) Room temperature steady-state absorption (black line) and photoluminescence (PL, red dashed line) spectra of CsPbBr₃ gp-QDs fabricated at 520 ℃. PL spectrum is measured with low-intensity 450 nm excitation. The second derivative of the absorption spectrum is shown by the blue line. The PL maximum and band-edge transition wavelengths (obtained from the minimum of the absorption second derivative) are indicated next to the respective curves; the measured (apparent) PL quantum yield (QY) is 1.2%. (b) The measured (red line) and the reconstructed (corrected for reabsorption; black dashed-dotted line) PL spectra of the same sample at room temperature. Inset is the results of accelerated photo-aging studies; the time axis shows the effective time under 1 sun illumination. These measurements were conducted for the sample fabricated at 490 ℃.

Using integrating-sphere measurements, we find that the apparent room-temperature PL QY of the gp-QD sample ($\eta_{PL}$) is ~1.2%. However, this value underestimates the actual QY because a significant overlap between absorption and PL spectra leads to reabsorption of a large fraction of the emitted light, especially on the blue side of the spectrum. This effect also leads to the distortion of the PL spectrum manifested in the asymmetry between its red and blue sides (Figure 1b, red line). To correct for reabsorption, we conduct a "reconstruction" of the true PL profile by fitting the red portion of the emission spectrum to the Voigt function, and then, extrapolating the fit to the blue side of the spectrum (black line in Figure 1b). The Voigt function represents a convolution of the Gaussian (*G*) and the Lorentzian (*L*) profiles and it can be expressed as (see Supplementary Note 1 of SM for the exact functional form of *G* and *L*):

$$V(h\nu, h\nu_0, \Gamma, \gamma) = A \cdot \int_{-\infty}^{\infty} G(h\nu', h\nu_0, \Gamma) L(h\nu - h\nu', \gamma) d(h\nu'), \qquad (1)$$

where *A* is the scaling factor (amplitude), $h\nu$ is the spectral energy, $h\nu_0$ and $\Gamma$ are the center and the width (standard deviation) of the Gaussian profile, respectively, and $2\gamma$ is the FWHM of the Lorentzian profile. This representation is well suited for describing QD samples, as in this case, the PL profile is determined by contributions from both inhomogeneous broadening due to distribution of QD sizes (typically Gaussian) and the homogeneous lineshape characteristic of individual QDs (typically Lorentzian). Later in this work, we use Equation 1 for the quantitative analysis of the temperature dependent PL broadening in our samples. Based on the corrected PL spectrum, the "true" PL QY of the gp-QD sample is ~1.7%, which is ~40% higher than the apparent efficiency. This, however, is still not an exact value but rather a lower limit as in our analysis we have assumed that reabsorption affects only the blue part of the emission spectrum, while in reality, its influence can extend past the PL center.

While being not very bright (at least at room temperature), the emission from these glass-based QD samples is extremely robust. According to the accelerated photostability tests, virtually no loss in the PL QY is observed after sample exposure to intense LED light with the dose equivalent to that of two years of natural sunlight illumination (Figure 1b, inset). This high level of photostablity is important for prospective applications of glass-based samples as, for example, LSCs.

Temperature-dependent PL measurements (Figure 2a) reveal a rapid growth of the emission intensity with decreasing *T*, which indicates increasing PL QY. At 220 K, $\eta_{PL}$ reaches ~5%, and then increases almost linearly (with *ca*. 0.1%/K rate) to

~25% at 20 K (Figure 2b). The mechanisms underlying this behavior are discussed in the next subsection where we analyze the results of time-resolved measurements.

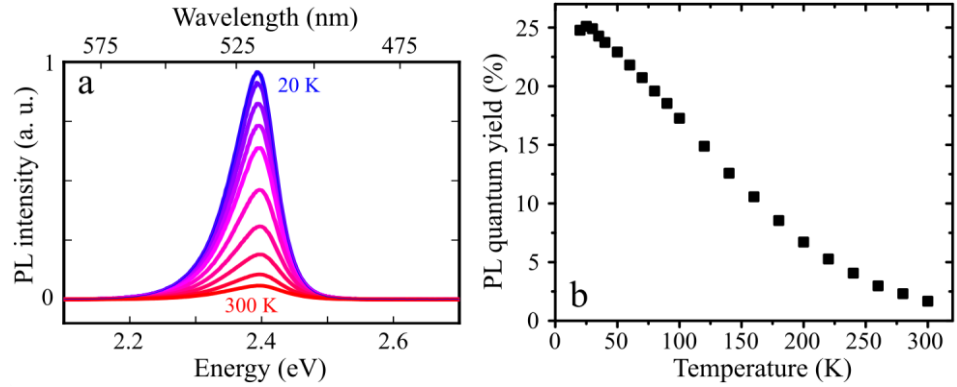

**Figure 2:** Temperature-dependent PL spectra and QY. (a) Temperature-dependent steady-state PL spectra of CsPbBr$_3$ gp-QDs ($T = 20 - 300$ K). (b) Temperature-dependent PL QY obtained by integrating the reconstructed steady-state PL spectra.

Simultaneously with increasing $\eta_{PL}$, we observe the narrowing of the PL spectrum (Figure 3a). To analyze it, we use a Voigt representation (Equation 1) and further assume that the inhomogeneous broadening (its Gaussian part) is temperature-independent, and hence, the observed linewidth changes are entirely due to variations in the homogeneous profile (i.e., the Lorentzian term). By applying a global fit procedure to PL spectra measured across the 20-to-300 K range (Figure 3a), we find $\Gamma = 120$ meV (note that the "true" linewidth is greater than the apparent value of 90 meV, as the latter is affected by reabsorption of the "blue" portion of the PL spectrum). We also obtain a set of $T$-dependent values of $\gamma$ displayed in Figure 3b. The homogeneous (i.e., single-dot) linewidth is typically linked to effects of exciton interactions with acoustic and longitudinal optical (LO) phonons and can be described by [32, 34-36]:

$$\gamma\left(T\right) = \gamma_0 + \sigma_a T + \frac{\Gamma_{LO}}{\exp\left(\frac{E_{LO}}{k_B T}\right)-1}, \qquad (2)$$

where $\gamma_0$ is the homogeneous broadening at $T = 0$ K, $\sigma_a$ is the exciton-acoustic phonon coupling coefficient, $E_{LO}$ and $\Gamma_{LO}$ are, respectively, the energy of the LO phonon and its characteristic line broadening parameter, and $k_B$ is the Boltzmann constant. Using Equation 2 with a fixed value of $E_{LO} = 18$ meV [37], we can closely fit the experimental $\gamma$-vs.-$T$ dependence (red line in Figure 3b), which yields $\gamma_0 = 18$ meV, $\sigma_a = 70$ μeV/K, and a $\Gamma_{LO} = 5$ meV. These values suggest that the dominant mechanism of homogeneous broadening is exciton-acoustic-phonon scattering. Previously, it was shown that in glass-embedded perovskite QDs, the mechanism of the line broadening strongly depends on the QD size. Specifically, in small-size QDs (~6 nm) it is the exciton-LO-phonon interaction, while in large-size QDs (~10 nm), it is the exciton-acoustic-phonon scattering [32]. In agreement with this previous assessment, in our ~11 nm QDs the dominant contribution to line broadening is provided by effects of exciton-acoustic-phonon scattering.

### 3.2 Temperature-dependent photoluminescence dynamics

To gain a deeper insight into the observed strong $T$-dependence of the measured PL QY (Figure 2a,b), we conduct measurements of time-dependent PL over the same range of temperatures (Figure 4a) using a low excitation fluence, which corresponds to the sub-single exciton regime when the average per-dot, per-pulse number of photoinjected electron-hole pairs $<N>$ is much less than unity ($<N> = $ ~0.05). The integration of the recorded PL transients yields values whose temperature dependence (blue triangles in Figure 4b) closely correlates with that of the steady state PL measured in the narrow spectral window around the wavelength used in transient PL measurements (open black squares in Figure 4b). This good correspondence suggests that the measured PL dynamics capture all time scales relevant to the emission process, and therefore, can be used to elucidate the mechanisms responsible for the observed T-dependence of the PL QYs.

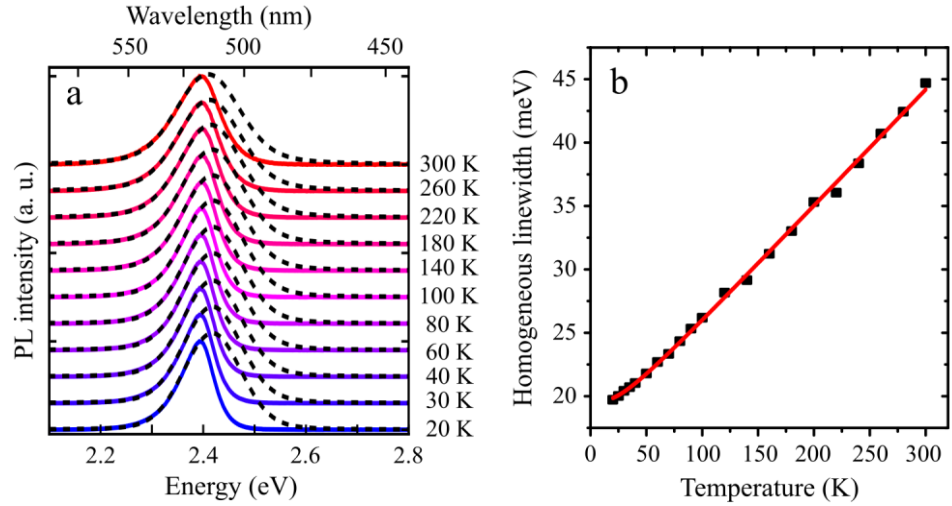

**Figure 3:** Temperature-dependent PL linewidth. (a) Temperature-dependent (($T = 20 - 300$ K), peak-normalized steady-state PL spectra of CsPbBr$_3$ gp-QDs. The measured spectra are shown by the colored solid lines. The spectra corrected for reabsorption are shown by the dashed black lines. (b) The dependence of the homogenous linewidth of the PL spectrum on temperature. Solid symbols represent the experimental data; the line represents the best fit to Equation 2 with fixed $E_{LO} = 18$ meV.

Previously, the observed drop in the PL QY of gp-QD samples with increasing $T$ has been explained in terms of thermally activated quenching described by the $T$-dependent nonradiative decay rate [32]. The inspection of time transients in Figure 4a immediately indicates that the varying rate model is incapable of explaining our experimental observations. We do observe the increase in the initial PL relaxation rate with increasing $T$; however, it has only a minor effect on the time-integrated signal, while the dominant role is played by strong changes in the early time PL amplitude. In fact, the comparison of the temperature dependences of the overall (time-integrated) PL signal (blue triangles in Figure 4b) and its amplitude (red circles in Figure 4b) indicates a very close correlation between the two, with a weak deviation at higher temperature ($T > 150$ K; see the ratio of the two quantities in the inset of Figure 4b), which is due to the effect of the varying relaxation rate, as discussed later in this work.

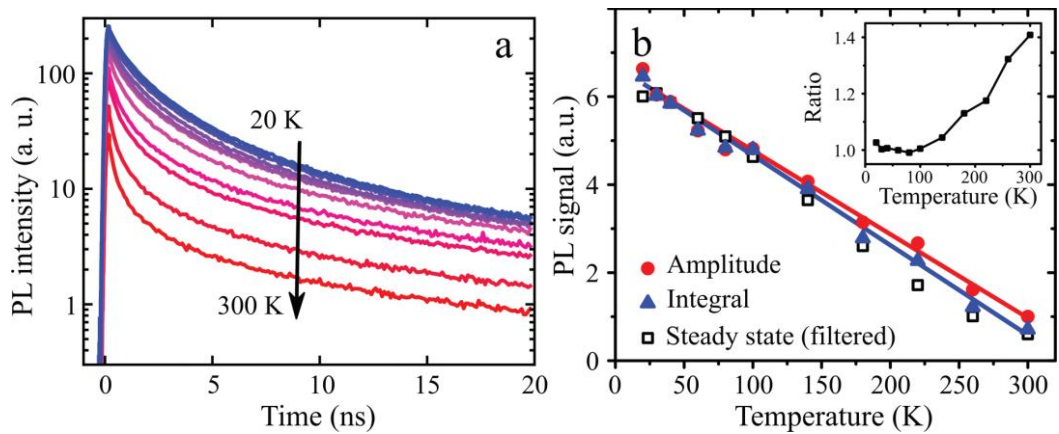

**Figure 4:** Temperature-dependent PL dynamics. (a) Temperature dependence of the time-resolved PL for 400-nm pulsed excitation (100 fs pump-pulse duration; $<N> \sim 0.05$); temporal resolution is ~70 ps. (b) Temperature dependences of the total PL intensity obtained by integrating the measured time transients (blue triangles) and the early time PL amplitude (red circles) along with linear fits (lines); open black squares show the temperature dependence of the spectrally filtered steady state PL signal measured around the wavelength at which the PL transients were recorded (540±20 nm). The inset shows the ratio of the early time PL amplitude and the time-integrated PL signal.

To explain this peculiar behavior of time resolved PL, we propose a model of two QD sub-ensembles, one of which contains emissive QDs, and the other, completely nonemissive particles. The large fraction of nonemissive dots was a typical property of early days glass-based II-VI QD samples and it was ascribed to very fast trapping at surface sites associated with unpassivated dangling bonds [38, 39]. As was observed in Ref. [38], these processes were characterized by extremely short times scales of the order of a few picoseconds which rendered the dots completely nonemissive. In perovskite QD samples, the sub-ensemble of "dark" dots can also be contributed by particles in nonemissive crystal phases whose fraction changes with temperature.

Based on the measured PL QY, in the studied samples, the nonemissive dots represent an absolute majority, and at room temperature, account for ~92% of all particles (see discussion later in this work). When the temperature is decreased, the fraction of these dots ($f_{nem} = 1 - f_{em}$; $f_{em}$ is the fraction of emissive dots) also progressively decreases, and at T = 20 K drops to less than 50%. The underlying reason for this behavior can be structural changes in QD surfaces that reduce the abundance of fast trapping sites, turning some of the nonemissive dots into emissive. An additional reason can be a transition between nonemissive and emissive crystal phases [21] initiated by the temperature change. An interesting peculiarity of the observed behavior is a nearly linear dependence of the $f_{em}$ on temperature (Figure 4b, red circles). This is again in contrast to expectations based on thermally activated nonradiative rate models, which normally predict a more complex dependence containing exponential term(s).

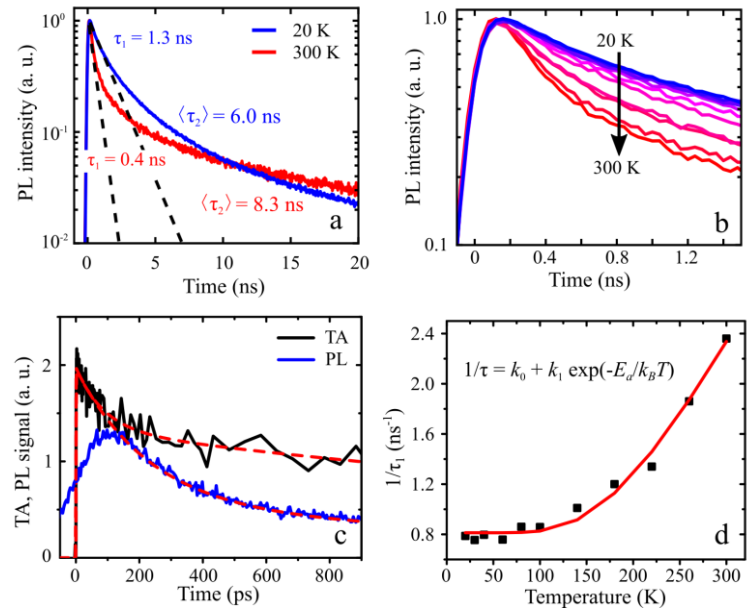

**Figure 5:** Temperature-dependent PL dynamics and comparison with TA. (a) Peak normalized PL dynamics measured at room temperature (red line) and at 20 K (blue line). The black dashed lines show single-exponential fits of the initial PL decay. (b) Peak-normalized early-time PL dynamics measured at different temperatures ($T = 20 - 300$ K). (c) Transient band-edge bleach (514-518 nm region; black) and PL dynamics recorded with excitation at 343 nm (TA) and 400 nm (PL); sub-single-exciton pump level with the average number of electron-hole pairs per dot per pulse <N> ~ 0.05. The resolutions of TA and PL measurements are, respectively, 100 fs and 70 ps. The delayed build-up of the PL signal is a result of the limited resolution. Therefore, when fitting the PL trace (red dashed line) we disregard it and instead extrapolate the fitting curve to $t = 0$. This allows for a more direct comparison of PL and TA dynamics. The fitting parameters are: $\tau_1 = 90$ ps, $A_1 = 33\%$, $\tau_2 = 3.1$ ns, $A_2 = 67\%$ for TA and $\tau_1 = 210$ ps, $A_1 = 70\%$, $\tau_2 = 2.2$ ns, $A_2 = 30\%$ for PL. (d) Temperature dependence of the initial PL decay rate (black squares). The red curve is the fit to the exponential dependence displayed in the panel.

While the major trend in the $T$-dependent PL QY can be ascribed to the $T$-dependence of $f_{em}$, there are more subtle variations in the emission intensity that can be understood based on details of slower relaxation processes captured in the measured PL dynamics. In Figure 5a, we compare peak-normalized PL transients recorded for $T = 300$ K (red line) and 20 K (blue line) using a low-pump-fluence (<N> ~ 0.05). An apparent distinction between these data is that the room-temperature trace is characterized by a faster initial decay than the low-temperature trace. At the same time, it exhibits a slower long-term dynamics. As a result, the two curves intersect at around 11 ns.

The initial decay time ($\tau_I$) found from a single-exponential fit of early time PL dynamics decreases from ~1.3 to 0.4 ns as $T$ changes from 20 to 300 K (Figure 5b). To answer the question whether this component can be assigned to radiative recombination, we compare the initial PL decay with the TA trace recorded under similar sub-single-exciton conditions (Figure 5c); see also Figure S3 of SM for a more complete set of TA data. In the case of radiative recombination, the population of the QD electron and hole states change at the same rate. As a result, the TA and PL should exhibit the same temporal behavior. This, however, is not the case in our measurements. Early-time PL decay is much more prominent compared to the TA relaxation indicating the existence of separate (independent) relaxation pathways for electrons and holes. This is a typical signature of a trapping process, when one carrier (an electron or a hole) is removed from the quantized state faster than the other. This immediately quenches PL, however, not TA as the TA signal is due to the additive contributions of electrons and holes, and therefore, can still remain nonzero even after one of the carriers leaves the QD band-edge state.

Based on above considerations, we assign the initial PL decay to nonradiative trapping. The measured $\tau_I$ constant indicates that the rate of this process increases from ~0.8 to 2.4 ns$^{-1}$ as $T$ is increased from 20 to 300 K. The observed temperature dependence (squares in Figure 5d) can be described in terms of thermal activation with the characteristic energy of ~61 meV, and the low-temperature rate $k_0 = 1/\tau_1(T=0) = 0.81$ ns$^{-1}$ (line in Figure 5d). Such thermally assisted trapping at surface/interface sites has been observed in many previous studies of QDs [35, 40, 41], and in particular, invoked to explain QD photoionization [42, 43].

Next we analyze the longer time PL component. It cannot be described by a single-exponential decay, therefore, we characterize it using a mean lifetime $<\tau_2>$ obtained by averaging the instantaneous time constant (defined as $I/(dI/dt)^{-1}$; $I$ is the PL intensity) over the time period which start from the point where the PL signal drops to 10% of its peak value and extends to 20 ns. Based on this definition, $<\tau_2>$ is 8.3 and 6.0 ns at $T = 300$ and 20 K, respectively. The room temperature $<\tau_2>$ constant is nearly the same as the previously measured radiative lifetime for colloidal CsPbBr$_3$ QDs of a similar size (*ca*. 7 – 8 ns) [33], suggesting that in the case of glass based samples the longer PL component also describes the radiative process. In fact, the observed slight shortening of $<\tau_2>$ at lower temperature is consistent with expectations based on the recent theory of the fine structure of band-edge excitonic states in these QDs [44]. According to this theory, the three closely spaced lowest-energy "emitting" states in CsPbX$_3$ dots are derived from a "bright" triplet while an optically passive "dark" singlet is located higher in energy. At cryogenic temperatures, the excitons reside in the lowest "bright" states, which corresponds to the high emission rate. At room temperature, a photoexcited exciton samples all four states including the nonemissive singlet. In the idealized scenario when all triplets have the same oscillator strength, while the singlet is totally "dark," the room-temperature emission rate would be 75% of the low-temperature rate. This is remarkably close to the ratio of the $<\tau_2>$ constants derived from the 300 and 20 K PL traces (6.0/8.3 = 0.72).

Using the above approximation for radiative lifetimes ($\tau_{rad} = <\tau_2>$), we can estimate the $T$-dependent PL QYs for the sub-ensemble of the emissive dots from the ratio of the overall average lifetime $<\tau>$ determined from the entire PL transient and $\tau_{rad}$. Using this procedure, we obtain that QY for the sub-ensemble of the emissive dots changes from ~43% to ~22% as the temperature increases from 20 to 300 K. This drop can be linked to the thermally activated nonradiative process analyzed in Figure 5d. The same process is responsible for the ~40% deviation between the $T$-dependences of the PL amplitude and the PL intensity time-integral at high temperatures (see Figure 4b and its inset).

Based on the measured steady-state PL QYs and the derived emission efficiencies within the sub-ensemble of emissive dots, we can estimate that the fraction of the nonemissive particles at $T = 20$ K is ~42%, *i.e.*, is less than half of the sample. However, it increases dramatically at room temperature, when $f_{nem}$ reaches ~92%. Thus, the primary goal of the efforts to improve room temperature emissivity of glass-based perovskite QDs is to understand the structural distinction between emissive and nonemissive particles and develop fabrication routines that would favor the QD structure for which the ultrafast trapping channel rendering dots nonemissive is shut off.

## 3.3 Biexciton dynamics

Previous studies of various types of colloidal QDs indicate a remarkable similarity (robustness) of multiexciton dynamics in samples with widely varied single-exciton relaxation behaviors [45, 46]. The multiexcion decay in these structures is dominated by a nonradiative Auger process whereby the electron-hole recombination energy is not emitted as a photon but instead dissipates as a kinetic energy of the third carrier (an electron or a hole) re-excited within the same band. In virtually all studied standard (not hetero-structured) QD systems, the Auger decay follows a universal volume scaling (or "V-scaling"), according to which the Auger lifetime is directly proportional to the QD volume.

Recent studies of Auger recombination in $CsPbX_3$ QDs also reveal extremely fast time scales of this processes that are even shorter than those in more traditional QD systems [33, 47]. These published works indicated the deviation of the measured Auger time constants from the standard *V*-scaling, which was attributed to the peculiarity of the confinement regime in this type of the QDs which was intermediate between strong and weak. These initial studies of multicarrier dynamics in perovskite QDs seem to indicate that the characteristics of Auger processes in these structures are not as universal as in the case of other studied QD materials. Therefore, it is unclear whether the Auger dynamics previously reported for colloidal perovskite QDs will be reproduced in glass-embedded samples. To clarify this question, here we conduct direct measurements of biexciton dynamics using pump-power-dependent TA.

In Figure 6a, we display TA traces recorded for <*N*> from 0.04 to 0.6. At the lowest fluences (0.04 - 0.08 excitons per dot), we observe self-similar dynamics that reflect single-exciton relaxation. When <*N*> reaches ~0.15, we observe the emergence of the fast initial component, which is a typical signature of biexciton Auger recombination. To isolate the biexciton dynamics, we subtract the tail-normalized single-exciton decay from the dynamics recorded for <*N*> = 0.15, 0.3, and 0.6, which produces a set of traces displayed in Figure 6b. All of them are characterized by the same exponential decay with a time constant of 50 ps, which represents the biexciton lifetime. This value is almost identical to one previously reported for colloidal $CsPbBr_3$ QDs of a similar size [33]. The similarity of biexciton dynamics is observed despite a dramatic difference in single-exciton decays between the glass-based and colloidal samples (Figure S4 of SM). These observations once again confirm a considerable robustness of Auger dynamics that are virtually insensitive to the quality of surface passivation or the type of the embedding medium.

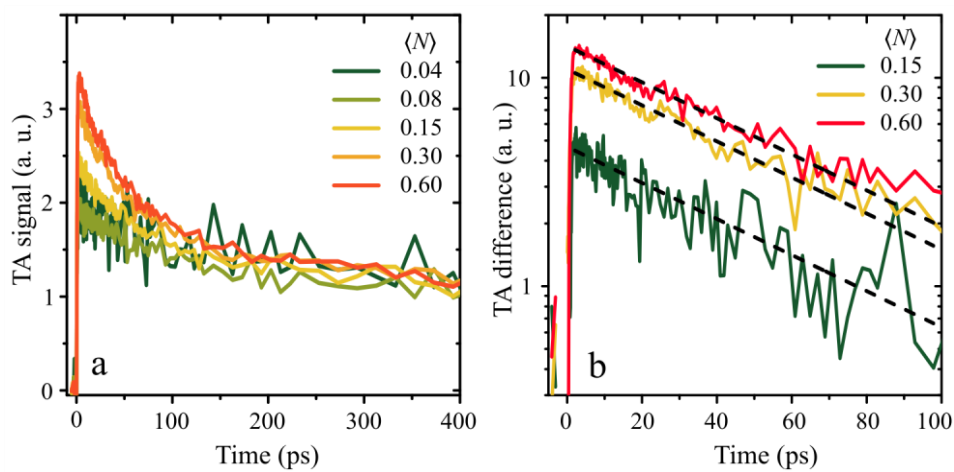

**Figure 6:** Biexciton Auger dynamics. (a) Pump-fluence dependent TA dynamics recorded with 343 nm excitation (100 fs pulse duration); the average number of excitons per QD is indicated in the legend. The traces are normalized at ~1 ns to highlight the difference in early-time dynamics due to the Auger recombination of photogenerated biexcitons. (b) The biexciton component of the TA decay isolated by subtracting the tail-normalized self-similar low-power dynamics (<*N*> = 0.04 - 0.08, black curve in Figure 5c) from the high-power decay dynamics (<*N*> = 0.15 - 0.6). The dashed lines show monoexponential fits that indicated the biextion lifetime of 50 ps.

# 4 Conclusions

We have investigated light emitting properties of a new type of $CsPbBr_3$ QDs prepared directly within a glass matrix via high temperature precipitation. The measurements of temperature-dependent PL indicate that a low temperature ($T = 20$ K) PL QY of these samples can be as high as ~25%. However, it quickly drops with increasing $T$, reaching 1−2% at room temperature. The PL intensity reduction closely correlates with the drop of the early-time PL amplitude but not the PL relaxation rate, and occurs following a peculiar, nearly linear dependence. This is in contrast to standard thermal quenching when the PL QY reduction directly follows the increase in the PL decay rate and typically shows an exponential temperature activated behavior. A possible explanation of these observations is the existence of a subset of nonemissive dots whose fraction is directly linked to the sample temperature. At $T = 20$ K, $f_{nem}$ is approximately 42%, and it increases to ~92% at $T = 300$K. The switching between the QD emissive and nonemissive state might occur as a result of a temperature-induced transformation in the surface/internal structure, which produces intragap states serving as very fast nonradiative recombination centers and/or nonemissive crystal phases [21, 48]. Detailed examination of transient PL and absorption reveals the presence of a temperature activated nonradiative process affecting the sub-ensemble of emissive dots. This process is characterized by sub-ns to ns dynamics and is responsible for ~50% drop of emissive sub-ensemble quantum yield. Finally, the high-pump fluence measurements reveal usual Auger decay signatures of biexciton states, with the relaxation time scale (~ 50 ps) being close to that previously observed for colloidal perovskite QDs [33].

## Acknowledgements

This work was supported by the Laboratory Directed Research and Development program at Los Alamos National Laboratory. We thank Maksym Kovalenko for insightful comments

# Supplementary Material

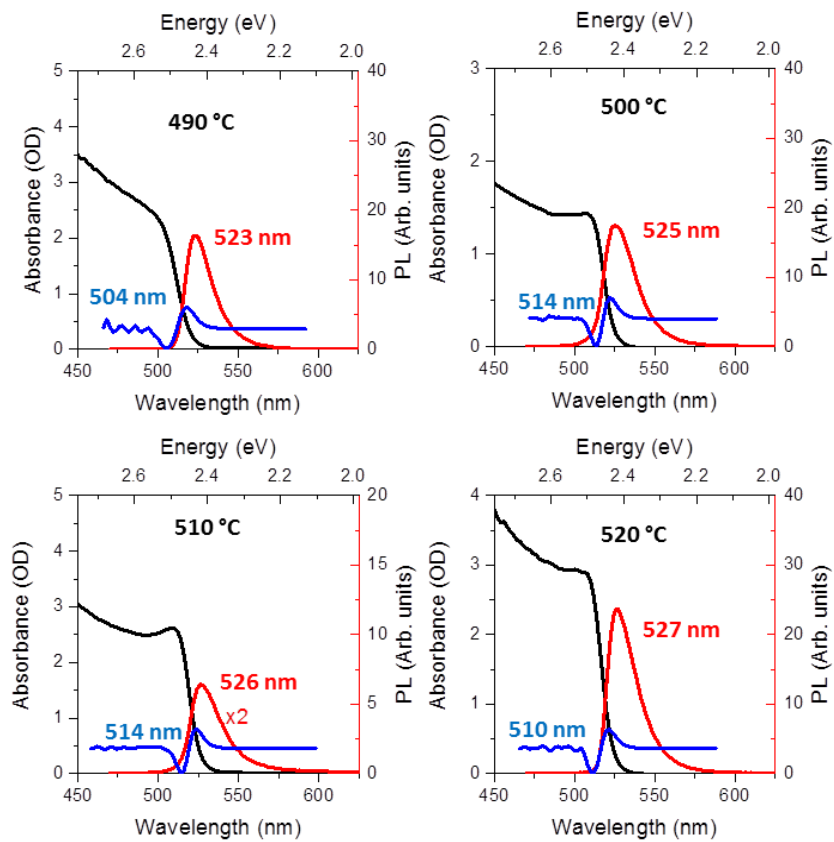

**Figure S1**: Room-temperature steady-state absorption (black lines) and photoluminescence (PL, red lines) spectra of glass-embedded CsPbBr$_3$ perovskite quantum dots (gp-QDs) fabricated at different temperatures (indicated in the figure). PL spectra are measured with 450 nm excitation. Second derivatives of the absorption spectra are shown by the blue lines. The PL maxima and band-edge transition wavelengths (obtained from the minima of the absorption second derivatives) are indicated next to the respective curves.

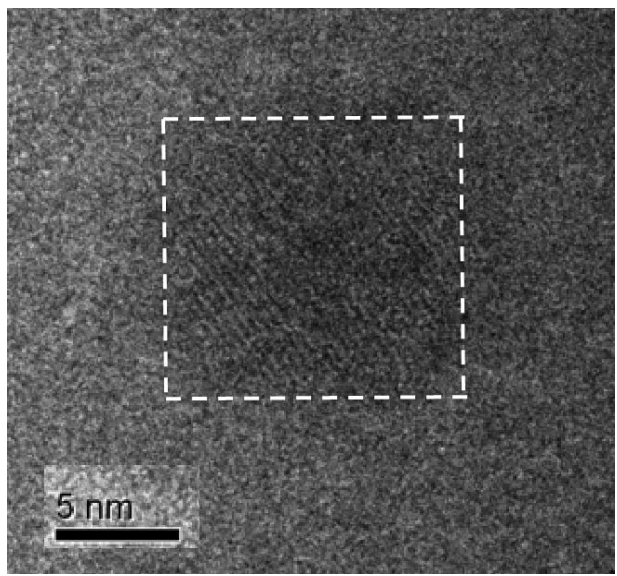

**Figure S2**: A high-resolution transmission electron microscopy (TEM) image of a glass-embedded CsPbBr$_3$ QD. The QD is approximately cubic with the side length of ~11 nm (white outline).

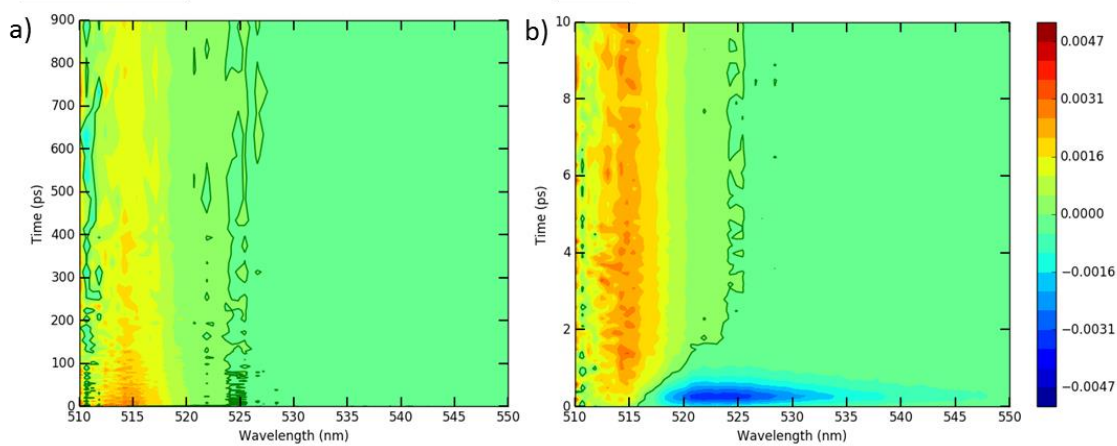

**Figure S3:** 2D contour plots of spectrally resolved transient absorption (TA) for the sample fabricated at 520 °C. Excitation wavelength is 343 nm and the average per-pulse, per-dot excitonic occupancy $\langle N \rangle$ is ~0.08. The contour line shows the zero of the TA signal. The full (0 – 900 ps) and the short (0 – 10 ps) ranges of TA data are shown, respectively, in (a) and (b). A prominent photobleaching in the 510-525 spectral range corresponds to the band-edge transition. The early time spectra (t < 1 ps) reveal photoinduced absorption below the band-edge bleach (520 – 530 nm). This is a signature of the transition shift due to a carrier-induced Stark effect. In the measurements shown in Figure 5c of the main text, the TA spectra were averaged over the 514-518 nm spectral range in order to avoid the interference from the Stark shift.

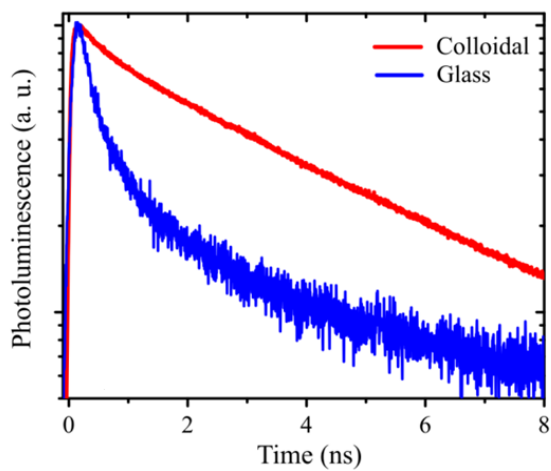

**Figure S1:** Comparison of peak-normalized time-resolved, room-temperature PL for glass-embedded (blue) and colloidal (red) CsPbBr$_3$ QDs ($<N> \sim$ 0.05). The faster PL decay in the glass-based sample is indicative of poorer surface passivation, which leads to enhanced nonradiative recombination via surface defects.

**Supplementary Note 1: Reconstruction of photoluminescence spectra using Voigt representation**

Strong reabsorption of glass-based perovskite QD samples distorts the shape of the measured PL spectra by selectively attenuating their blue side (see, e.g., Figure 1a,b of the main text). To account for this effect, we fit the red part of the PL spectrum to the Voigt function and then use a "mirror image" of this fit to describe the blue part of the spectrum. The Voigt function can be presented as:

$$V(hv, hv_0, \Gamma, \gamma) = A \cdot \int_{-\infty}^{\infty} G(hv', hv_0, \Gamma) L(hv - hv', \gamma) d(hv'),$$

where A is the amplitude, $hv$ is the spectral energy,

$$G(hv, hv_0, \Gamma) = \frac{1}{\sqrt{2\pi\Gamma^2}} e^{-\frac{(hv - hv_0)^2}{2\Gamma^2}}$$

is a Gaussian profile centered around $hv_0$ with the width (standard deviation) $\Gamma$, and

$$L(hv, \gamma) = \frac{\gamma}{\pi((hv)^2 + \gamma^2)}$$

is a Lorentzian profile with FWHM = $2\gamma$.